\documentclass[dvips]{PoS}

\title{Chiral symmetry and lattice fermions}

\ShortTitle{Chiral symmetry and lattice fermions}

\author{\speaker{Michael Creutz} \thanks { This manuscript has been
    authored under contract number DE-AC02-98CH10886 with the
    U.S.~Department of Energy.  Accordingly, the U.S. Government
    retains a non-exclusive, royalty-free license to publish or
    reproduce the published form of this contribution, or allow others
    to do so, for U.S.~Government purposes.  I'm greatful for my
    Alexander von Humboldt Research Award which partially supported my
    attendence to this meeting.  }\\ Brookhaven National
  Laboratory\\ E-mail: \email{creutz@bnl.gov}}

\abstract{Lattice gauge theory and chiral perturbation theory are
  among the primary tools for understanding non-perturbative aspects
  of QCD.  I review several subtle and sometimes controversial issues
  that arise when combining these techniques.  Among these are one
  failure of partially quenched chiral perturbation theory when the
  valence quarks become lighter than the average sea quark mass and a
  potential ambiguity in comparisons of perturbative and lattice
  properties of non-degenerate quarks.}

\FullConference{From quarks and gluons to hadronic matter: A bridge
  too far?\\ 2-6 September, 2013\\ European Centre for Theoretical
  Studies in Nuclear Physics and Related Areas (ECT*), Villazzano,
  Trento (Italy)}

\begin{document}

\section{Introduction}




\def \psibar{\overline\psi}

As is well known, understanding QCD at low energies requires a
non-perturbative treatment.  While several tools exist, here I will
concentrate on two of the most successful, chiral symmetry and lattice
gauge theory.  Combining these two has a tortuous history, marked with
several bitter controversies that continue to this day.  In this
talk I will fan these fires further.

In the next section I review some rather straightforward predictions
of chiral symmetry for the meson spectrum in two flavor QCD with
non-degenerate quarks.  I restrict myself to two flavors for
simplicity, none of the basic points depend on this in any essential
way.  Everything in Section \ref{chiral} is well known and I believe
not controversial in any way.  Then in Section \ref{dashen} I become
less conventional and explore what would happen if the up quark mass
were to become negative while maintaining the down quark mass at a
positive value.  At a certain point the neutral pion can become
massless, and beyond that one should expect a pion condensation into
what is known as the Dashen phase \cite{Dashen:1970et}.

Section \ref{messages} takes the picture from the previous sections
and draws a few general messages which indicate non-trivial
consequences for a Euclidean path integral, normally the starting
point for lattice gauge simulations.  In particular, I will point out
some of the peculiarities appearing in the structure of the fermionic
operators entering these path integrals.  Section \ref{lattice} I
discuss how a generic lattice fermion action automatically has a sort
of chiral symmetry.  This symmetry is associated with the requirement
of doublers, which are pushed to high energy with Wilson or overlap
fermions.  Section \ref{summary} summarizes the main messages and some
of the controversial consequences thereof.

\section{Chiral symmetry and two flavor QCD}
\label{chiral}

Throughout this discussion I will concentrate on two flavor QCD with
non-degenerate quarks, labeled as usual as up $(u)$ and down $(d)$.
The restriction to two flavors is for simplicity, and the basic ideas
go through for more species.  This section is meant to remind you of
some of the standard consequences of chiral symmetry for the light
pseudo-scalar masses.  In this theory one can construct four fermion
bilinears that transform as pseudo-scalars
\begin{equation}
\matrix{
i\overline u\gamma_5 d\sim\pi_+
&&i\overline d\gamma_5 u\sim\pi_-
&&i\overline u\gamma_5 u
&&i\overline d\gamma_5 d.\cr
}
\end{equation}
The first two of these create the charged pions.  Usual chiral
symmetry arguments give these mesons squared masses proportional to
the average of the quark masses
\begin{equation}
M^2_{\pi_\pm}\sim (m_u+m_d)/2.
\end{equation}

The other two pseudo-scalar combinations, both electrically neutral,
conceal some rather interesting issues.  Gauge theories generically
conserve helicity in their interactions.  Separating the helicity
combinations of these operators naively suggests that the mixing of
\begin{equation}
\overline u\gamma_5 u
=\overline u_L u_R-\overline u_R u_L
\end{equation}
with
\begin{equation}
\overline d\gamma_5 d
=\overline d_L d_R-\overline d_R d_L
\end{equation}
would be suppressed by the product of the up and down quark masses.
This would further suggest that there are two light neutral
pseudo-scalars, one with $M^2_{\overline u\gamma_5 u}\sim m_u$ and a
second with $M^2_{\overline d\gamma_5 d}\sim m_d$.

Of course it is well known that this prediction is wrong.  The chiral
anomaly strongly mixes the operators $\overline u\gamma_5 u$ and
$\overline d\gamma_5 d$.  This mixing occurs through what is sometimes
called the effective ``t'Hooft vertex'' \cite{'tHooft:1976fv}, which
for this problem takes the form
\begin{equation}
\overline u_L u_R \overline d_L d_R
+\overline u_R u_L \overline d_R d_L.
\end{equation}

\begin{figure}
\begin{centering}
\includegraphics[width=.5\textwidth]{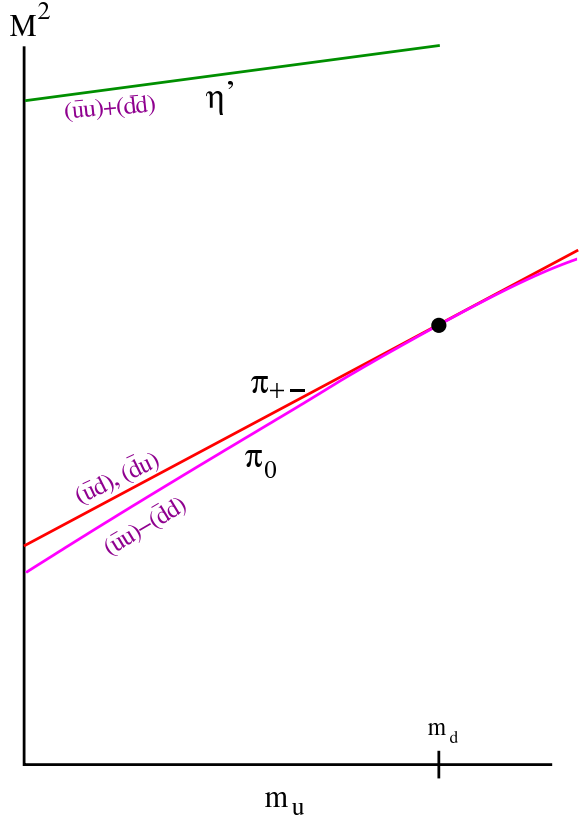}
  \caption{A schematic representation of the pseudo-scalar mass
  dependence on the up quark mass with a fixed non-vanishing down
  quark mass.  The theory maintains a mass gap even at vanishing up
  quark mass.}
\label{iso1}
\end{centering}
\end{figure}

This mixing breaks naive chiral symmetry and has dramatic consequences
for the pseudo-scalar spectrum.  In particular, the symmetric combination
\begin{equation}
\eta^\prime\sim i\overline u\gamma_5 u + i\overline d\gamma_5 d
\end{equation}
is not a pseudo-Goldstone boson.  Its mass does not become small in
the chiral limit, but rather remains of order the QCD scale
\begin{equation}
M_{\eta^\prime}\sim \Lambda_{qcd}+O(m_u+m_d).
\end{equation}
After this mixing the orthogonal combination remains to represent the
neutral pion
\begin{equation}
\pi_0 \sim i\overline u\gamma_5 u - i\overline d\gamma_5 d.
\end{equation}
Since this is made up of half up and half down quarks, the neutral and
charged pions are degenerate up to quadratic order in the quark masses
\begin{equation}
\label{splitting}
M_{\pi_0}^2=M_{\pi_\pm}^2-O((m_u-m_d)^2)-\hbox{e.m. effects}.
\end{equation}
Of course, in the physical world the electromagnetic effects dominate
this mass difference, but here I ignore this for simplicity.

As these points are crucial to the later discussion, it is perhaps
useful to sketch the picture graphically.  Consider fixing the down
quark mass and study the meson spectrum as the up quark mass is
varied.  As sketched in Fig. \ref{iso1}, the three pions have a
leading mass dependence going as $M_\pi^2 \propto {m_u+m_d\over
  2}+O(m_q^2)$ while the eta prime is heavier.  The effect of isospin
breaking is to give a small separation of the neutral and charged
pions as in Eq.~(\ref{splitting}).  The sign of this splitting is not
determined by symmetry alone, but it is natural to expect the neutral
pion to be lighter because of mixing with the heavier eta prime and
any pseudo-scalar gluon bound states.  One important observation from
this picture is that a mass gap persists even when the up quark is
massless.

\section{The Dashen phase}
\label{dashen}

\begin{figure}
\begin{centering}
\includegraphics[width=.6\textwidth]{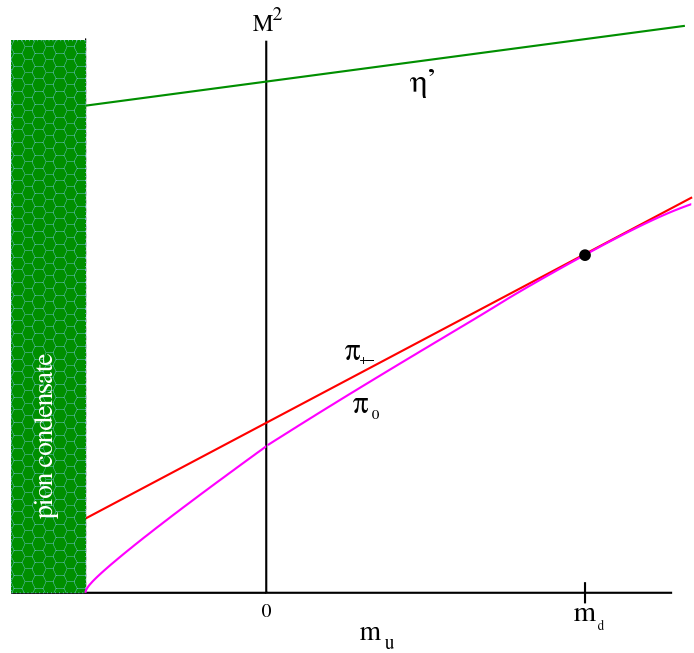}
\caption{As the up quark mass is varied into the negative mass region,
  the pions will continue to get lighter until the neutral pion mass
  vanishes.  Beyond that point one expects a CP violating phase where
  the neutral pion field has an expectation value.}
\label{iso2}
\end{centering}
\end{figure}

The existence of a mass gap at $m_u=0$ as shown in Fig. \ref{iso1}
suggests that there is no physical singularity as the up quark mass is
further varied into the negative region.  Such an extrapolation is
expected to behave smoothly up to a point where the square of the
neutral pion mass goes to zero.  Beyond that, and this occurs sigma
models, one should expect the neutral pion will spontaneously condense
and acquire an expectation value.  Since this is a CP odd field, the
resulting phase will spontaneously break CP.  The resulting extension
of Fig. \ref{iso1} is sketched in Fig. \ref{iso2}.

This possibility of a spontaneous breaking of CP in the strong
interactions was suggested in 1971 by Dashen \cite{Dashen:1970et}.
This was before the understanding that QCD was the likely underlying
theory; Dashen's discussion was based on the current algebra ideas of
the day.  This behavior is also natural in sigma models for chiral
symmetry breaking, both in the linear \cite{Creutz:1995wf} and the
non-linear \cite{Creutz:2003xu} forms.  Those models also suggest that
the transition is Ising like, with the order parameter being the
expectation of the neutral pion field.  This transition formally
occurs where the strong CP angle $\Theta$ takes the value $\pi$; it
occurs in a region where the product of the quark masses is negative.
The $\Theta$ parameter as usually defined is the phase of this
product.

\begin{figure}
\begin{centering}
\includegraphics[width=.5\textwidth]{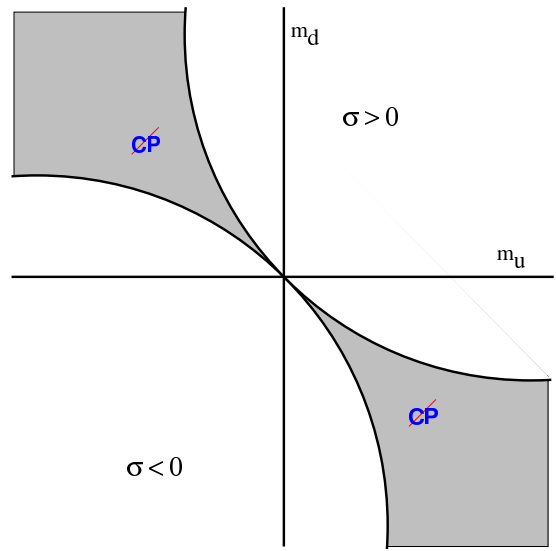}
\caption{Varying both the up and down quark masses, a CP violating
  phase lies in a portion of the region where the product of these
  masses is negative.}
\label{iso4}
\end{centering}
\end{figure}

Fig. \ref{iso4} shows another way of looking at this phase structure.
On varying the two physical parameters $m_u$ and $m_d$, a CP violating
phase covering a portion of the region where their product is
negative.  The symmetries of this figure are quite instructive.  In
particular the picture is symmetric in the sign of the average quark
mass ${m_d+m_u\over 2}$ as well as in the quark mass difference
${m_d-m_u\over 2}$.  However there is no symmetry between these two
quantities.  While the vanishing of either of these quantities is
protected by the symmetry, non-perturbative effects are expected to
renormalize them differently.

\section{Messages and consequences for the path integral}
\label{messages}

The picture presented in the previous sections is quite simple and
should not be particularly controversial.  However when this physics
is thought about in terms of the path integral approach, it raises
several interesting points.  These are at the heart of several of the
controversies dividing the lattice community.

\subsection
{Message 1: {A mass gap persists when only one quark is massless
}}

This point is connected to several issues.  For one, it means that QCD
with only one massless quark has no infrared issues.  Of course
perturbative QCD requires care with infrared divergences, but these
are an artifact of perturbation theory.  For any gauge invariant
correlator, long distance physics will fall exponentially in the
distance with the mass of the lightest state that can be exchanged;
that is,
\begin{equation}
\langle \phi(x)\phi(0)\rangle_c\ <\sim\ e^{-m_{\pi_0} |x|}.
\end{equation}
This may be a bit unexpected for the path integral.  While there can
in principle be small Dirac eigenvalues in the quark matrix, no long
distance physics arises.  Of course this is special to the one flavor
theory; with 2 or more massless quarks the pions become massless.

Note that the prediction from the sigma models does not show any
non-analyticity as the up quark mass passes through zero.  This raises
the question of whether there is any experimental signature of a
vanishing up quark mass.  Most discussions assume that this point is
well defined, but this has never been proven non-perturbatively.  The
fact that the average quark mass and the quark mass difference have no
symmetry between them indicates that there is no symmetry to protect
the point of vanishing up quark mass when the down quark mass doesn't
vanish.  More formally, it is unclear whether different
non-perturbative regularization schemes, such as the use of different
overlap operators, will have a universal continuum limit for the
theory at vanishing up quark mass.

The absence of a singularity at $m_u=0$ has further consequences for
the spectrum of the Dirac operator appearing in the path integral.
Banks and Casher \cite{Banks:1979yr} long ago pointed out a relation
between the density of small Dirac eigenvalues and the jump in the
chiral condensate that occurs on passing through the chiral point.
Since the non-degenerate quark system being explored here has no jump
in physics in the $m_u=0$ region, there cannot be a finite density of
small eigenvalues for the up quark operator.

This point raises serious issues for a popular technique known as
``partially quenched chiral perturbation theory.''  This approach
considers adding ``valence'' quarks to the theory which have no
feedback on the gauge fields.  The physical dynamical quarks are then
referred to as ``sea'' quarks.  The usual assumption is that multiple
valence quarks will undergo a chiral condensation and give rise to
massless valence pions as their mass goes to zero.  It is known that
this approximation fails in the pure gauge theory without dynamical
quarks \cite{Svetitsky:2005qa,Golterman:2005fe,Golterman:2004cy}.
With a massless dynamical quark, however, one might expect the theory
to be better behaved.  Nevertheless, as a valence quark mass
approaches the up quark mass, the sea and valence propagators become
identical.  If the sea quark mass does not have small eigenvalues,
then the valence quarks cannot either.  Thus the usually assumed
chiral condensation cannot occur if only the up sea quark is massless.
The basic conclusion is that the partially quenched theory cannot be
trusted for
\begin{equation}
m_{valence} < \langle m_{sea} \rangle.
\end{equation}

\subsection{Message 2: {The sign of a quark mass is physically relevant}}

It should be immediately clear from Fig.~\ref{iso2} that the meson
spectrum is different between a small positive up quark mass and a
negative value of the same magnitude.  This is a crucial property of
QCD that cannot be seen in perturbation theory.  Indeed, in any
perturbative diagram the sign of the mass appearing in the quark
propagators can be reversed by a chiral rotation.  This rotation,
however, is anomalous.  

The reason the sign of the mass becomes relevant is most easily
understood as arising from gauge configurations of non-trivial
topology.  In particular, when one mass is negative, the path integral
receives a weighting factor of $(-1)^\nu$, where $\nu$ is the winding
number of the background gauge configuration.  This is a special case
of the more general result that if the up quark mass is given a phase
$m_u\rightarrow e^{i\Theta\gamma_5}m_u$ then the theory describes the
so called ``Theta vacuum.''  This is a physically different theory for
each $\Theta$.  I will return to the effects of this rotation later in
the discussion.

The relevance of the sign of the quark mass has important consequences
for attempts to relate lattice parameters to those obtained from
perturbation theory in the continuum.  This means that attempts to
match lattice and $\overline{MS}$ masses are particularly dangerous
for non-degenerate quarks.

\subsection
{Message 3: {A divergent correlation length is possible
when no quarks are massless}}
 
Fig.~\ref{iso2} indicates that at the point where the
Dashen phase begins the neutral pion mass will vanish.  In the
effective chiral models this is an Ising like second order transition.
This occurs at a point where neither the up nor down quark masses
vanish, although they are of opposite sign.

In terms of the Dirac operators in the path integral, this shows that 
the mass gap can vanish without having any small Dirac eigenvalues.
The location of this transition is a dynamical issue, requiring the
tuning of the up quark mass to the edge of the Dashen phase.

At a somewhat deeper level, this picture is connected to whether or
not there is a first order transition at $\Theta=\pi$.  When the
lightest quarks are multiply degenerate, it is fairly straightforward
to argue that there must be a first order transition at such a point
\cite{Creutz:2009kx}.  But in the case where the lightest quark is
non-degenerate, there is a transition only once one enters the Dashen
phase.  In that phase as one crosses $\Theta=\pi$, the order parameter
$\langle\pi_0\rangle$ undergoes a jump.  But before one reaches this
phase, that is when the up quark mass is negative but small, the
behavior at $\Theta=\pi$ is smooth.  Thus whether there is a
transition or not is a delicate dynamical issue.  The corresponding
question of a possible transition at $\Theta=\pi$ in the zero flavor
theory remains open.

Another interesting feature of the boundary of the Dashen phase is
that at this point the topological susceptibility diverges to minus
infinity \cite{Creutz:2013xfa}.  This comes about since the operator
$F_{\mu\nu}\tilde F_{\mu\nu}$ is expected to have a finite amplitude
to create a neutral pion.  This gives an infrared divergence to the
susceptibility as the neutral pion mass goes to zero.  Reflection
positivity \cite{Seiler:2001je} forces this divergence to be negative.

\section{Lattice Fermions}
\label{lattice}

I now turn to some generic properties of lattice Dirac operators and
the connection to the chiral anomaly.  These points are closely
related to the Nielsen-Ninomiya theorem \cite{Nielsen:1980rz}, but my
approach will be a bit different.

Suppose I am given some arbitrary lattice Dirac operator $D$ to use
for a lattice simulation.  To proceed, I assume gamma five hermiticity
\begin{equation}
 \gamma_5 D \gamma_5=D^\dagger.
\end{equation}
This is a property of the majority of the operators currently used in
practice.  (One notable exception is the twisted mass approach which adds a
chiral rotation to this property.)  

Now consider dividing the operator $D$ into Hermitean and
antihermitean parts $D=K+M$ with
\begin{equation}
\matrix{
&K=(D-D^\dagger)/2\cr
&M=(D+D^\dagger)/2\cr
}
\end{equation}
This immediately implies
\begin{equation}
\matrix{
[K,\gamma_5]_+=0\cr
[M,\gamma_5]_-=0.\cr
}
\end{equation}
On a lattice everything is fully regulated and finite; so the naive
equation ${\rm Tr}\gamma_5=0$ still holds.  An immediate consequence
is that
\begin{equation}
M\rightarrow e^{i\theta\gamma_5}M
\end{equation}
is an exact symmetry of the
determinant.  Explicitly I have
\begin{equation}
|K+M|=|e^{i\gamma_5 \theta/2}(K+M)e^{i\gamma_5 \theta/2}|
=|K+e^{i\theta\gamma_5}M|
\end{equation}
This leads to the next message.

\subsection{
Message 4: Any lattice action is symmetric under the chiral rotation
$M\rightarrow e^{i\theta\gamma_5}M$}

This looks rather peculiar in the context of the earlier remark that a
chiral rotation of the mass term yields an inequivalent theory with a
non-vanishing $\Theta$ parameter.  Indeed, where did the anomaly hide?
The answer is that this particular symmetry can only be a flavored
chiral symmetry, which is allowed by the anomaly.  But for this to be
a flavored symmetry, all lattice actions must bring in some extra
structure.  

Some actions, such as naive, staggered, and minimally doubled fermions
resolve this by having doublers.  Half of them use $\gamma_5$ and half
$-\gamma_5$ for their chiral rotations.  In this way the naive chiral
symmetry above is actually a flavored symmetry.

\begin{figure}
\begin{centering}
\includegraphics[width=.45\textwidth]{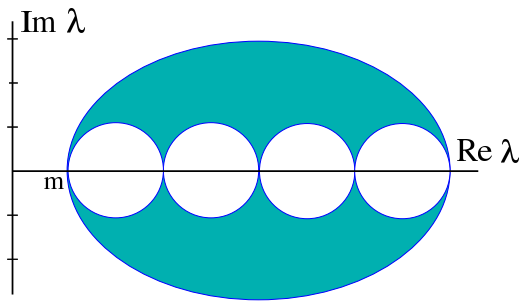}
\caption{The eigenvalue spectrum of the free Wilson fermion operator
  forms a set of nested ellipses.}
\label{eigen1}
\end{centering}
\end{figure}

Wilson and overlap fermions contain this symmetry in a more subtle
manner.  In these cases the Hermitean part $M$ ceases to be a
constant.  Instead, it produces heavy states that cancel the anomaly.
To see this more explicitly, consider Wilson fermions.  For the free
theory the Dirac operator takes the momentum space form
\begin{equation}
 D_W  =
{1\over a} \sum_\mu(i\gamma_\mu\sin(p_\mu a)+1-\cos(p_\mu a)) 
+m.
\end{equation} 
The corresponding eigenvalue structure takes the form of a set of
nested ellipses and is sketched in Fig.~\ref{eigen1}. 

For small momentum the ``mass'' term takes the form
\begin{equation}
m+{1\over a}(1-\cos(p_\mu a))=m+O(a)
\end{equation}
however for momentum components near $\pi$ the eigenvalues are of
order $1/a$.  Note that as seen in Fig.~\ref{eigen1}, there is no
symmetry if $m$ changes sign.  As noted earlier, this is a necessary
feature of QCD, and with Wilson fermions it is manifest from the
outset.  This action also breaks naive chiral symmetry explicitly,
{\sl i.e.} $[D_w,\gamma_5]_+ \ne 0$ even when $m=0$.

\begin{figure}
\begin{centering}
\includegraphics[width=.8\textwidth]{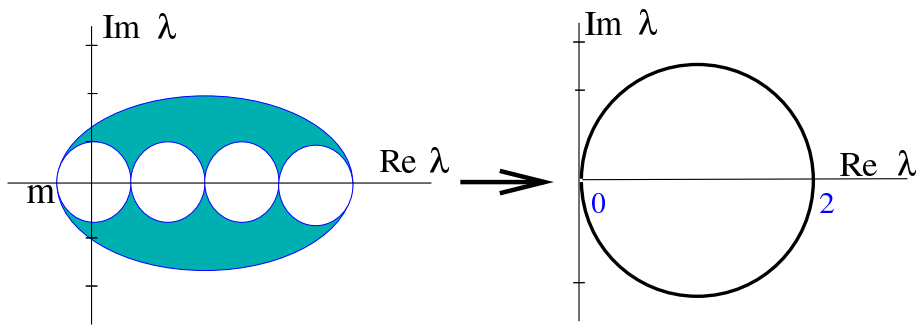}
\caption{The overlap operator is obtained by projecting 
  eigenvalues of the Wilson fermion operator
  onto a unit circle.}
\label{eigen3}
\end{centering}
\end{figure}

The overlap operator \cite{Neuberger:1997fp} is obtained by projecting
the Wilson operator onto a circle to obtain a unitary matrix $V$ which
is then shifted to give small eigenvalues near the origin.
\begin{equation}
D_W\rightarrow D_o=1-V, \qquad V^\dagger V=1.
\end{equation}
This
process is sketched in Fig.~\ref{eigen3}.

This projection process is computationally demanding, but it leaves an
operator with rather elegant properties.  First, it maintains gamma
five hermiticity $\gamma_5 D_o \gamma_5=D_o^\dagger$.  Second, unlike
in the Wilson case, it is a normal operator; {\sl i.e.} it commutes
with its adjoint, $[D_o^\dagger,D_o]=0$.  But probably most important,
it retains an exact variant of chiral symmetry
\begin{equation}
e^{i\theta\hat\gamma_5} D_o e^{i\theta\gamma_5} = D_o
\end{equation}
where 
\begin{equation}
\hat\gamma_5 =V\gamma_5.
\end{equation}
As with $\gamma_5$, $\hat\gamma_5$ is Hermitean and $\hat\gamma^2=1$.
All its eigenvalues are $\pm 1$
and allow one to define an index.
\begin{equation}
\nu={1\over 2}{\rm Tr}\hat\gamma_5={\rm Tr}{\gamma_5+\hat\gamma_5\over 2}.
\end{equation}
This definition agrees with the continuum index for smooth fields.

With these preliminaries, it should now be clear where the anomaly
went with Wilson and overlap fermions.  Effectively the doublers are
still there, but have been given masses of order the cutoff.  The
chiral rotation $M\rightarrow e^{i\theta\gamma_5}M$ not only rotates
the light quark masses, but also those of the heavy doubler states.
This emphasizes an old observation of Seiler and Stamatescu
\cite{Seiler:1981jf} that the physical $\Theta$ is the relative angle
occurring under independent rotations the fermion mass and the Wilson
term.

With the overlap operator, any zero eigenmodes corresponding to
topology will have heavy counterparts on the opposite side of the
unitary circle.  Rotating the Hermitean part of this operator rotates
these heavy modes as well. 
 
This hiding of the anomaly is not just a lattice artifact, but is
relevant to the continuum theory as well.  In general the physical
$\Theta$ can be moved around and placed on any flavor at will.  In the
standard model, any non-zero $\Theta$ can be entirely moved into the
top quark phase.  Since the value of $\Theta$ is physical, {\sl i.e.}
it gives a neutron an electric dipole moment, some properties of the
top quark remain relevant to low energy physics!  In other words, the
naive decoupling theorems for heavy quarks do not apply
non-perturbatively.

\section{Summary}
\label{summary}

The main messages of this talk are:
\begin{enumerate}
\item A mass gap persists when only one quark is massless.
\item The sign of a quark mass is physically relevant.
\item A divergent correlation length can occur even
when no quarks are massless.
\item Any lattice action is symmetric under the chiral rotation
$M\rightarrow e^{i\theta\gamma_5}M$.
\end{enumerate}
These are all rather straightforward consequences of effective chiral
Lagrangians, but are less obvious when thought of in terms of fermion
determinants in a Euclidean path integral.  Some of the more
controversial consequences are
\begin{enumerate}
\item There is no proof of universality between non-perturbative
  schemes for the point where the up quark mass vanishes when the
  other quarks are massive.
\item The usual assumptions of partially quenched chiral perturbation
  theory can fail if $m_{valence} < m_{sea}$.
\item Matching lattice results to {$\overline{MS}$} can miss
non-perturbative effects when the quarks are non-degenerate.
\item Decoupling theorems do not apply non-perturbatively.
\end{enumerate}

There are other related controversial points that I have not covered
in this talk.  One is that the topological susceptibility also has an
ambiguity from rough gauge fields.  This is closely tied to the above
non-universality of the point for vanishing up quark mass.  Finally
there is the infamous rooting issue for staggered quarks.  In the
previous editions of this conference
\cite{Creutz:2009zq,Creutz:2011ej} I explained in some detail as to
why that approach is incorrect; I will not go into it further here.



\end{document}